\documentstyle[12pt]{article}

\begin{document}

\topmargin -12mm
\oddsidemargin 5mm

\setcounter{page}{1}
\vspace{1cm}
\begin{center}

\bf IMPROVED CRYSTAL METHOD FOR PHOTON BEAM LINEAR POLARIZATION
MEASUREMENT AT HIGH ENERGIES\\
\vspace{3mm}
{\large A.B. Apyan\footnote {Corresponding author. E-mail:
aapyan@inix.yerphi.am},
 R.O. Avakian, S.M. Darbinian, K.A. Ispirian, S.P. Taroian }\\
Yerevan Physics Institute, Yerevan, Armenia
\end{center}

\begin{abstract} 
A method for photon linear polarization determination based on the
measurement of the asymmetry of pairs produced by polarized photons in
single crystals within the optimal intervals of pair particles energies is
proposed. In difference to the well known methods the asymmetry in this
case is essentially larger. The optimal orientation of crystal is found
which provides the maximal values for analyzing power and figure of merit
as well as minimal measurement time.
\end{abstract}

\indent

\vspace{9mm}  
\centerline{{\bf{1.Introduction}}}

\indent
At present there is a demand for production of linearly and still more of 
circularly polarized photons with energies $ \omega \geq 20 $ GeV
because after the 
beautiful series of the SLAC [1] and few others photoproduction experiments 
carried out with linearly and circularly polarized photon beams at $ \omega 
\leq 10 $ GeV no published experimental 
data exist besides those obtained at CERN SPS on $ \phi $- photoproduction 
with the help of linearly polarized photons with $ \omega $ = 20-75 GeV [2]. 
The SLAC beam was produced by Compton scattering of polarized laser photons 
on SLAC, 20 GeV electrons, and there was no necessity to measure its high
polarization P $ \approx 90 \% $, while the polarization $ P \approx 30 \%
$ of the CERN beam produced by coherent bremsstrahlung on Si single
crystal has been measured [3] with the help of the asymmetry of the decay
$ \rho^0 \rightarrow \pi^+ + \pi^- $ because some experimental factors
could influence and change the expected value of P. Now the operation of 
these as well as of another polarized photon beam used for few QED
experiments only [4] is stopped.

\indent
On the other hand there are many proposed experiments (see [5]) devoted to 
QCD and nucleon form factors problems which can be 
realized with the help of polarized photon beams with $ \omega \geq 20 $ GeV 
produced at SLAC and Fermilab as it is described in [6,7] and [8], 
respectively. In this connection the measurement of the photon beam 
polarization becomes important. Unfortunately, 
the methods of polarization measurements at $ \omega \leq 10 $ GeV (see 
[9]) are not applicable at higher energies and low intensities. The
$\rho$-decay asymmetry method used in [3] is connected with difficulties,
due to the relatively low yield of $\rho$ mesons (about $ 10^{-5} $ per
photon,
see [5]) and the necessary $\pi / e $  angular discrimination to eliminate
the high intensity e$^{+}$ e$^{-} $-background.

\indent
Taking into account the above said and the perspectives of the realization of 
the polarization conversion with linear and circular polarization measurements 
by the $\rho$-decay method according to the proposal [5] it is necessary
to revise the possibilities of the methods using single crystal. Let us
note that as the preliminary measurements [10] have shown one of the
recently proposed method [11] measuring the asymmetry of total pair yield
with y = $\varepsilon_{\pm}/\omega $ = 0 -1 ($ \varepsilon_{\pm}$ is the
energy of produced pair particles) without magnetic spectrometer has not
provided practical results at CERN SPS because of the presence of low
energy photons accompanying high energy photons with multiplicity 1.5-3 in
the beam [12].

\indent
In this work to escape the difficulties connected with the photon beam 
multiplicity it is suggested to use magnetic spectrometer, and since the 
method is for relatively low intensity beams it is proposed to use
asymmetric pairs of the coherent pair production in a much wider 
y-interval in contrast to the well known method [13] when one uses only 
symmetric pairs with  y $ \approx $ 0.5  within a narrow y-interval with 
$ \Delta$ y/ y $\approx 1- 2 \%$. The method can be tested and 
used in the experiment [5] devoted to the conversion of the linear 
polarization into circular in a way described in [5,14]. Though diamond
crystals provide better results all the calculations are carried out for 
silicon single crystals because large diamond crystals covering the
electron beam cross section are not available and are expensive. 

\vspace{9mm}
\centerline{\bf{2. The Physics and Formulae of the Method }}

\indent
To illustrate the physics of the proposed method Fig.1 shows the 
y-dependence of the differential absorption coefficients
of the coherent pair production in Si single crystal by 100 GeV photons
in case of "point effect" orientation when polar entrance angle with
respect to the axis [110] is $ \theta = 40$ mrad and the azimutal angle 
between the plane $(1\bar{1}0)$ and the plane containing the photon
momentum $\vec{k}$ and axis [110] is  $ \alpha = 21.8 $ mrad. As it
follows from Fig.1 the differential analyzing power
\begin{equation}
\label{AA}
\rho = \mathrm\frac{ dw_{\parallel}/{dy}-dw_{\perp}/{dy}}
          {dw_{\parallel}/{dy}+dw_{\perp}/{dy}},\
\end{equation}
where dw$_{\parallel,\perp}$/dy are the differential pair absorption
coefficients with photon polarization vector $\vec{\mathrm{e}}$ parallel 
and perpendicular to the plane $(\vec{\mathrm{k}},[110])$, is sufficiently
high and approximately constant within a wide y-interval 
\begin{equation}
\label{AB}
\mathrm{y_1 \leq y \leq y_2}  .
\end{equation}
Here $ \mathrm{y}_{1,2} $ are the values of y when the differential
absorption coefficients $ \mathrm{dw_{\parallel, \perp }/dy} $ have a
jump. The calculations show that the value of the asymmetry for choosed 
pairs within the interval (2) appears to be significantly larger than the 
asymmetry for pairs within the interval y=0-1. Therefore in this work we 
propose to use the experimental asymmetry
\begin{equation}
\label{AC}
\mathrm {a =\frac{n_{\parallel} - n_{\perp}}{n_{\parallel} +
n_{\perp}}}
\end{equation}
with the numbers of pairs  n$_{\parallel ,\perp }$
produced in the interval (2). The degree of linear
polarization is determined by the formula P=a/r where

\begin{equation}
\label{AD}
\mathrm{r = \frac{w_{\parallel} - w_{\perp }}{w_{\parallel} + w_{\perp
}}},
\end{equation}
and w$_{\parallel , \perp } $ are the theoretical integral absorption 
coefficients, i.e. the integrals of the $ \mathrm{dw_{\parallel , \perp
}/dy} $ over the interval (2).
In difference from the notations of [11] W, R and F calculated
for the full interval (0,1) of y in this paper the pair production 
characteristics integrated over the interval (2) are denoted by small 
letters w, r and f for absorption coefficients, asymmetry and figure of 
merit, respectively.
\indent
The values of $\mathrm{y_{1,2}} $ for given reciprocal lattice vector
$\vec{\mathrm{g}}$ are  determined from the condition  g$_{\parallel} =
\delta$
\begin{equation}
\label{AE}
\mathrm{y_{1,2}(g) = \frac{1}{2}(1 \pm \sqrt{1-\tau(g)})},
\end{equation}
where g$_{\parallel}$ is the component of $\mathrm {\vec{g}}$
parallel to the photon momentum  $\vec{\mathrm{k}}$,
$\delta=\mathrm{m^{2}/2}\omega \mathrm{y(1-y)}$,
$\delta_{\mathrm{m}}=\mathrm{2m^{2}/}\omega$ is the minimal momentum
transfer and 
$\tau(\mathrm{g})=\delta_{\mathrm{m}}/\mathrm{g}_{\parallel}$, 
$\mathrm{(\hbar=c=1)}$. In the case of\quad "point effect" when the
reciprocal 
lattice vector (220) gives the main contribution we have $\mathrm{ \tau(g)
= bm/4 \sqrt{2} \pi \lambda_{c} \omega \theta sin \alpha }$ ( b is the
lattice constant, $ \lambda_{\mathrm{c}} $ is the electron Compton 
wavelength). If the momentum acceptance $\mathrm{(y_{min}}$, 
$\mathrm{y_{max})}$ of the pair spectrometer is not sufficiently large 
only pair components with $ \mathrm{y_{min}\ge y_{1}}$, and/or 
$\mathrm{y_{max}\le y_{2}}$ can be detected, hence the limits of 
integrations of the differential absorption coefficients $\mathrm{y_{1}}$
and/or $\mathrm{y_{2}}$ should be replaced by $\mathrm{y_{min}}$ and/or 
$\mathrm{y_{max}}$. In the following calculations we assume that the pair 
spectrometer acceptance is large and all pairs within the interval (2) can 
be accepted.

\indent
Following [11] we present the polarization ($\mathrm{\vec{e}}$) dependent 
integral absorption coefficients w in the form
\begin{equation}
\label{AF}
\mathrm{w(\omega, \vec{e}) = w^{am} + w_{1}^{coh}(\omega) +
w_{2}^{coh}(\omega, \vec{e})}. 
\end{equation}
The amorphous part is independent of $\mathrm{ \vec{e}} $ and is equal to
\begin{equation}
\label{AG}
\mathrm{w^{am} = w_{0}{[B_{1}(y_{2}) - B_{1}(y_{1})]\psi_{1} -
[B_{2}(y_{2}) - B_{2}(y_{1})]},}
\end{equation}
where in the exponential and Moliere screening approximation
\begin{equation}
\label{AH}
\mathrm{\psi^{exp}_{1}=4[0.5+ln(m/\beta)-B(s)],}
\end{equation}
\begin{equation}
\label{AI}
\mathrm{\psi^{Mol}_{1}=4[0.5+ln(111Z^{-1/3})-
\sum_{i=1}^{3} a^{2}_{i}B(s_{i})\-
\sum_{i\ne j}^{3} a_{i}a_{j}\frac{b^{2}_{j}}{b^{2}_{i}-b^{2}_{j}}
exp(s_{i})E_{i}(s_{i})].}
\end{equation}
Here $\mathrm{\beta^{-1}=R}$ is screening radius, $\mathrm{u_{1}}$ is mean
square thermal oscillation amplitude, $\mathrm{a_{i}}$ and
$\mathrm{b_{i}}$ are parameters of the Moliere potential, B(x) is a well
known function [15] of $\mathrm{s=\beta^{2}u_{1}^{2}}$ and
$\mathrm{s_{i}=\beta_{i}^{2}u_{1}^{2}}$, 
$\mathrm{\beta_{i}=mb_{i}Z^{1/3}/121}$ and
Ei(x) is the exponential integral. The functions $\mathrm{B_{1,2}}$ are 
given by the expressions
\begin{equation}
\label{AJ}
\mathrm{B_{1}(x)=x-2x^{2}/3+4/3x^{3}/9,\quad
B_{2}(x)=2(x^{2}-2x^{3}/3)/9.}
\end{equation} 
The first of the coherent parts depends on $\omega$ and is equal to
\begin{eqnarray}
\label{AK}
\mathrm{w^{coh}_{1}}(\omega)&=&\mathrm{w_{0}\delta_{m}\sum_{\vec{g}}
D(\vec{g})\frac{g_{\perp}^{2}}{g_{\parallel}^{2}}
\{[A_{1}(y_{2})-A_{1}(y_{1})]} \nonumber \\
&&\mathrm{+\tau(g)[A_{2}(y_{2})-
A_{2}(y_{1})]-\tau^{2}(g)[A_{3}(y_{2})-A_{3}(y_{1})]/4}\
\end{eqnarray}
and the second coherent part which depends on $\mathrm{\omega }$ and
$\mathrm {\vec{e}}$ is equal to
\begin{equation}
\label{AL}
\mathrm w^{coh}_2(\omega, \vec{e}) = w_{0}\delta_m\sum_{\vec{g}}
D(\vec{g})
\frac{g_{\perp}^{2}}{g_{\parallel}^{2}}\frac{\tau^{2}(g)}{8}
[1 - 2 |\hat{g}_{\perp}\vec{e}|^{2}][A_{3}(y_{2})-A_{3}(y_{1})],
\end{equation}
where
\begin{equation}
\label{AM}
\mathrm{D(\vec{g})=\frac{2\pi^{2}}{V_{0}}\frac{S(\vec{g})}{n_{0}}
exp(-g^{2}u_{1}^{2})
\left[\frac{1-F(\vec{g})}{g^{2}}\right]^{2},}
\end{equation}
\begin{equation}
\label{AN}
\mathrm A_{1}(x) = ln\frac{x}{1-x} - 2x,\quad
A_{2}(x) = ln\frac{x}{1-x},\quad
A_3(x) = \frac{1}{1-x} - \frac{1}{x} + ln\frac{x}{1-x}.
\end{equation}
In the above expressions $\mathrm{ w_{0} = n \sigma_{0}}$, $\mathrm{
\sigma_{0} = Z^{2} r_{0}^{2}/137 }$, n and Z  are the atomic density and
nucleus charge of the crystal, $ \mathrm{r_{0}} $ is classical electron 
radius, m is electron mass, $ \mathrm{\vec{g}_{\perp}}$ is the component 
of $\mathrm{\vec{g}} $ perpendicular to $\mathrm{\vec{k} }$, 
$\mathrm{\hat{g_{\perp }} =  \vec{g_{\perp }}/g_{\perp}} $. In the 
formulae for $\mathrm{w^{coh}_{1,2}}$ the summations are carried out for 
all vectors $\mathrm{\vec{g}}$ with $\mathrm{ g_{\parallel}\ge 
\delta_{m}}$ and the limits of integration $\mathrm{y_{1,2}}$ must be
replaced by (5) if $\mathrm{y_{1,2}(g)}$ are less than $\mathrm{y_{1,2}}$.
In the case of integration over full interval (0,1) substituting
$\mathrm{y_{1}=0}$ and $\mathrm{y_{2}=1}$ in $\mathrm{w^{am}}$ and
replacing $\mathrm{y_{1}}$ and $\mathrm{y_{2}}$ by (5) we obtain the total 
absorption coefficients $\mathrm{W^{am}}$ and $\mathrm{W^{coh}}$ and 
corresponding R=($\mathrm{W_{\parallel} - W_{\perp })/(W_{\parallel } + 
W_{\perp} }$).

\indent
In conclusion of this section let us write the expression for the  number of 
primary photons $\mathrm{ n_{\gamma}} $\quad necessary to measure the beam 
linear polarization P with an accuracy $\mathrm{ \Delta P} $ [11].
\begin{equation}
\label{AO}
\mathrm n_{\gamma} = 1/(tf(\Delta P)^2)
\end{equation}
where t is the crystal thickness in cm and
$\mathrm{f=r^{2}(w_{\parallel}+w_{\perp})}$ is the figure of merit.
 
\vspace{9mm}
\centerline{\bf{3. Numerical Results and the Measurement of the Linear 
Polarization }}

\indent
To obtain numerical results with the help of the formulae (6-15) first we
calculate the dependence of $\mathrm{ w_{\parallel, \perp}} $ upon
$\mathrm{\theta_{y} = \theta
sin \alpha }$ (see Fig.2 for a [110] Si crystal for $ \mathrm{\omega =
100} $ GeV ). The variation of the angle $\mathrm{ \theta_{x} = \theta cos
\alpha }$ in a wide interval does not change the results. Then we
calculate the figure of merit f, entering in (15), depended on the
angle $\mathrm{\theta_{y}}$ ( see Fig. 3 for $ \mathrm{\omega = 100 }$
GeV). The optimal orientation angle $\mathrm{ \theta^{opt}_y }$
corresponding to the maximal value of the figure of merit
$\mathrm{f^{max}}$ gives the orientation for wich minimal measurement time
is required for the given accuracy $ \mathrm{\Delta P/P }$ of the
polarization measurement.

\indent
Fig.4 shows the energy dependencies of the optimal $\mathrm
{\theta^{opt}_y}$ and $\mathrm{ r^{opt}}$ corresponding to the maximal
figure of merit $\mathrm{f^{max}}$. For comparison in Fig.4 is shown also
optimal $\mathrm{ R^{opt} }$, calculated by the use 
$\mathrm{W^{opt}_{\parallel,\perp}}$\quad[11]. In this case the asymmetry
is low $ \mathrm{R^{opt}\approx 20\%}$ in comparison to the 
$\mathrm{r^{opt}\approx 40\%}$ for photon energy $\mathrm{\omega=100 }$
GeV and at the same crystal orientation ($\mathrm{\theta_{x}=40 }$mrad,
$\mathrm{\theta_{y}=0.873}$mrad). As the calculations show the optimal 
angles $\mathrm{ \theta^{opt}_y }$ corresponding to the $\mathrm{f^{max}}$
and the $\mathrm{F^{max}}$ coincide. The $ \omega $-dependencies
of\quad$\mathrm{f^{max}}$ and corresponding $\mathrm{w^{opt}_{\parallel,
\perp}} $ as well as $ \mathrm{w^{am}} $  are shown in Fig.5. For
comparison in Fig.5 is also shown y dependence of\quad$ \mathrm{F^{max}}
$. Having the results of the Figs.4 and 5 one can choose the optimal
parameters of the crystal polarimeter. Indeed, coming out from the photon
energy and the energy interval in the spectral distribution one firstly
chooses the crystal angle $\mathrm{ \theta^{opt}_y} $ using the
corresponding curve of Fig.4. Then using the value $\mathrm{r^{opt}}$ from
Fig.4 and the values of $\mathrm{ w^{opt}_{\parallel, \perp}} $ from Fig.5
and giving the desired measurement accuracy $\mathrm{ \Delta P/P }$ one
finds the number of necessary photons for the given thickness  of crystal
with the help of formula (15). The choice of t is conditioned by the
compromise between available thickness, fast gathering of necessary
statistics and allowed radiation energy losses of the produced pair
particles in the crystal polarimeter. Determining the polarization with
the help of the formula P=a/r one must use the asymmetry measured only for
pairs with y values in the interval (2) and (5) for the given photon
energy and orientation.

\vspace{9mm}

\centerline{{\bf{4. Summary}}}

\indent
The obtained formulae (6)-(15) and numerical results allow to use the
described method for the measurement of polarization when the photon beam
intensity is low as in the case of experiment [5]. The large value of
the figure of merit and the reduced measurement time is the advantage of
the proposed method.\\
This work was supported by ISTC Grant A099.

\newpage
\centerline{{\bf{Figure Captions}}}

Fig.1 The differential absorption coefficients $\mathrm{dw_{\parallel} /
dy} $ (solid line) and $\mathrm{ dw_{\perp}/dy} $ (dashed line) as a
function of y for [110] Si at $ \mathrm{\omega = 100} $ GeV. The incidence 
angles are $\mathrm{\theta_{x} }$= 40 mrad and $\mathrm{ \theta_{y}} 
$=0.873 mrad ($\mathrm{y_{1,2}=0.37, 0.63}$).\\
\indent
Fig.2 The absorption coefficients $\mathrm{w_{\parallel}} $ and $
\mathrm{w_{\perp} }$ and
$\mathrm{w=(w_{\parallel}+w_{\perp})/2}$ 
as a function of $\mathrm{ \theta_{y}} $ for [110] Si at 
$\mathrm{\omega=100 }$ GeV.\\
\indent
Fig.3 The figure of merit $\mathrm{f = r^{2}(w_{\parallel}+w_{\perp}) }$
as a function of $\mathrm{\theta_{y}} $ for [110] Si at $\mathrm{
\omega =
100} $ GeV ($\mathrm{\theta_{x} = 40}$ mrad).\\
\indent
Fig.4 The energy dependencies of $\mathrm{ \theta^{opt}_y }$,
$\mathrm{r^{opt}}$ and $\mathrm{R^{opt}}$ calculated with the use of
absorption coefficients $\mathrm{ w^{opt}_{\parallel,\perp}} $ and $
\mathrm{W^{opt}_{\parallel,\perp}} $,\quad respectively.\\
\indent
Fig.5 The energy dependencies 
of\quad$\mathrm{f^{max}}$, $\mathrm{F^{max}}$, 
$\mathrm{w^{opt}_{\parallel,\perp }}$ and $\mathrm{ w^{am}}$.

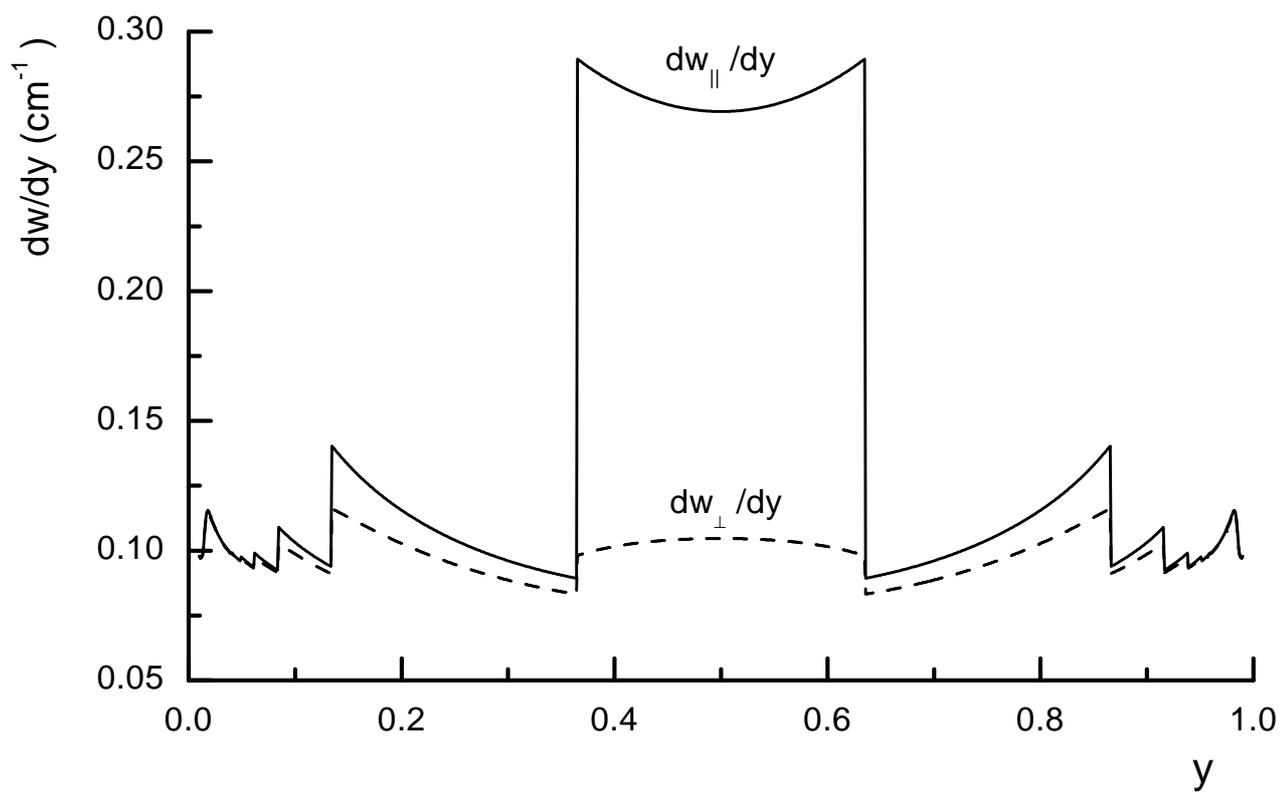
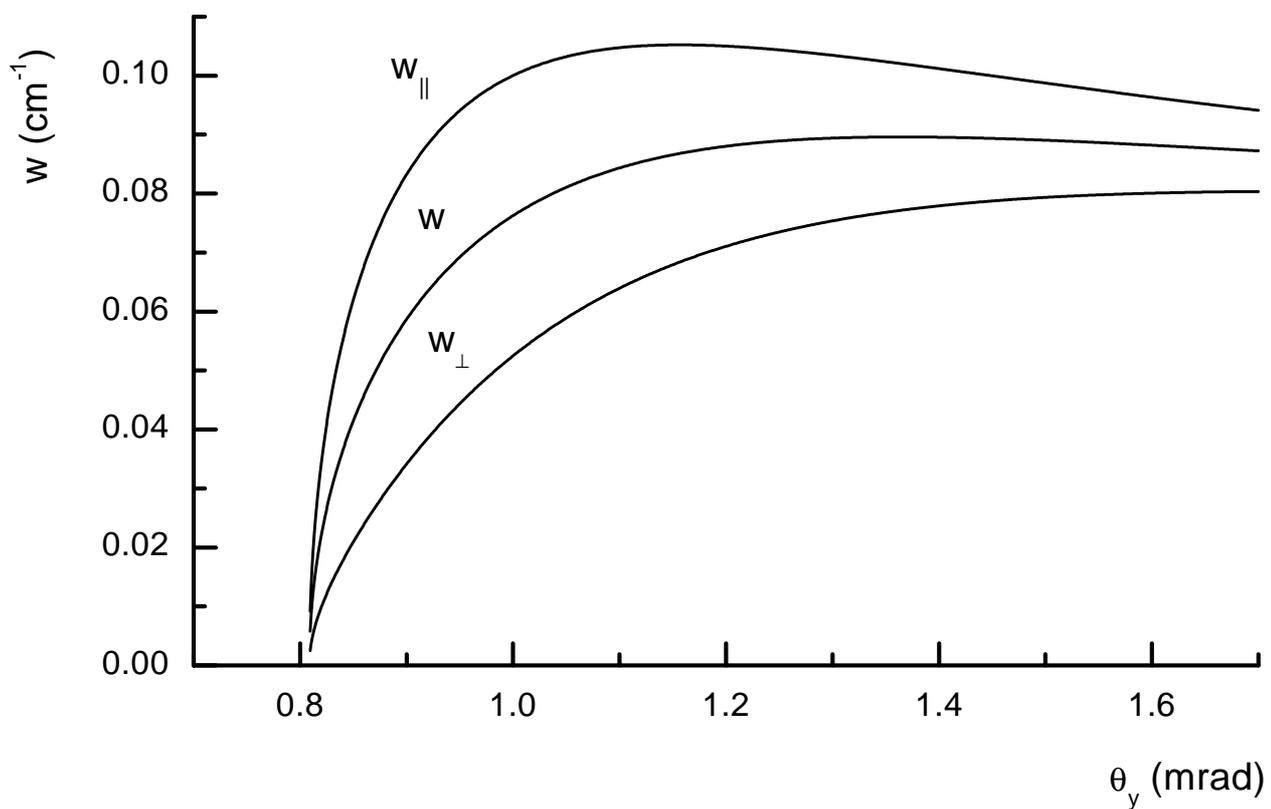
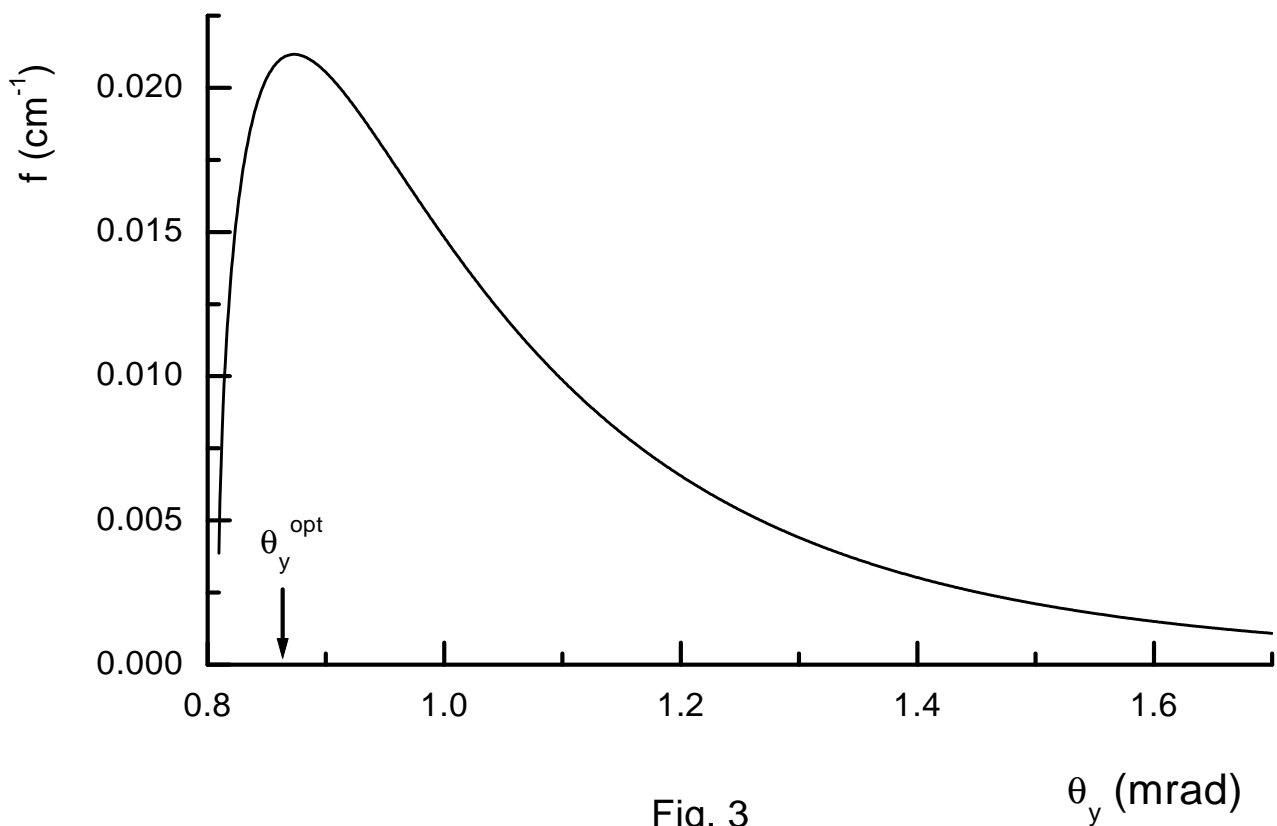
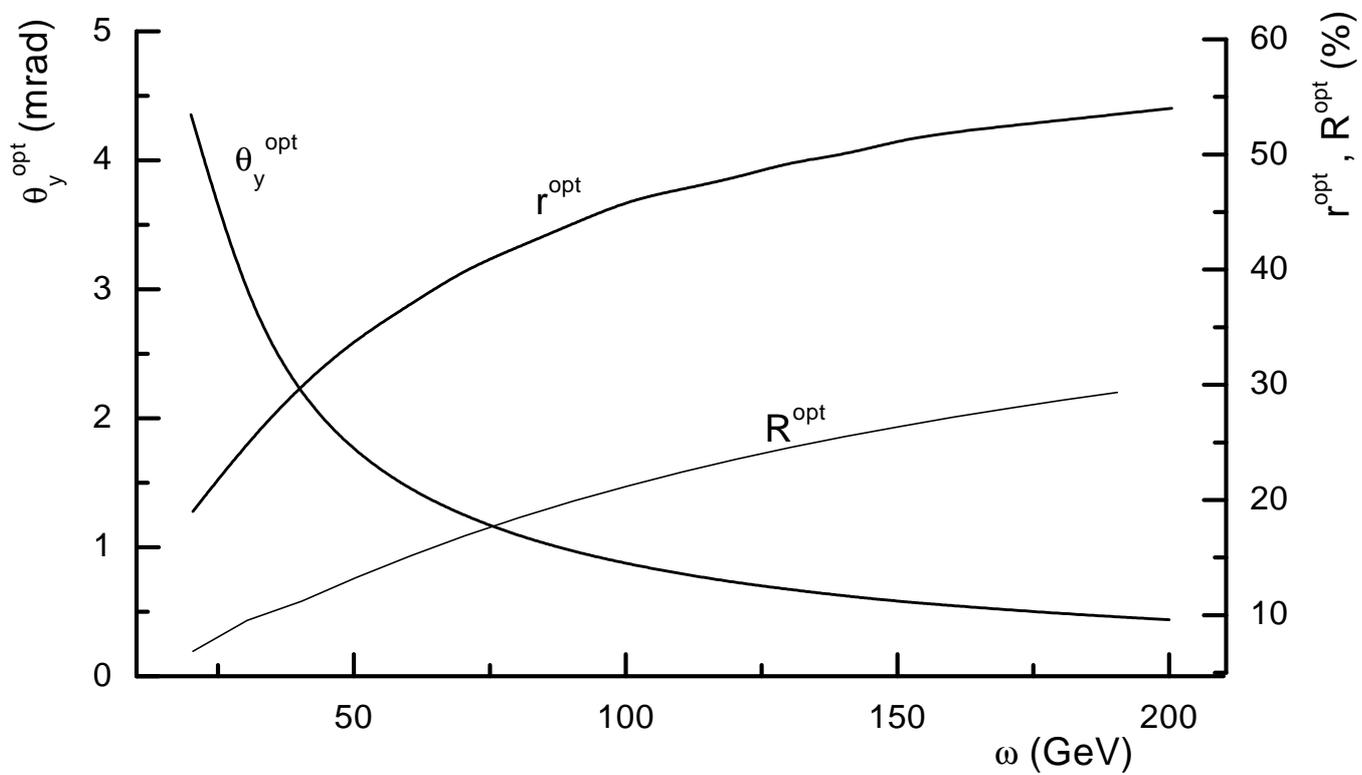
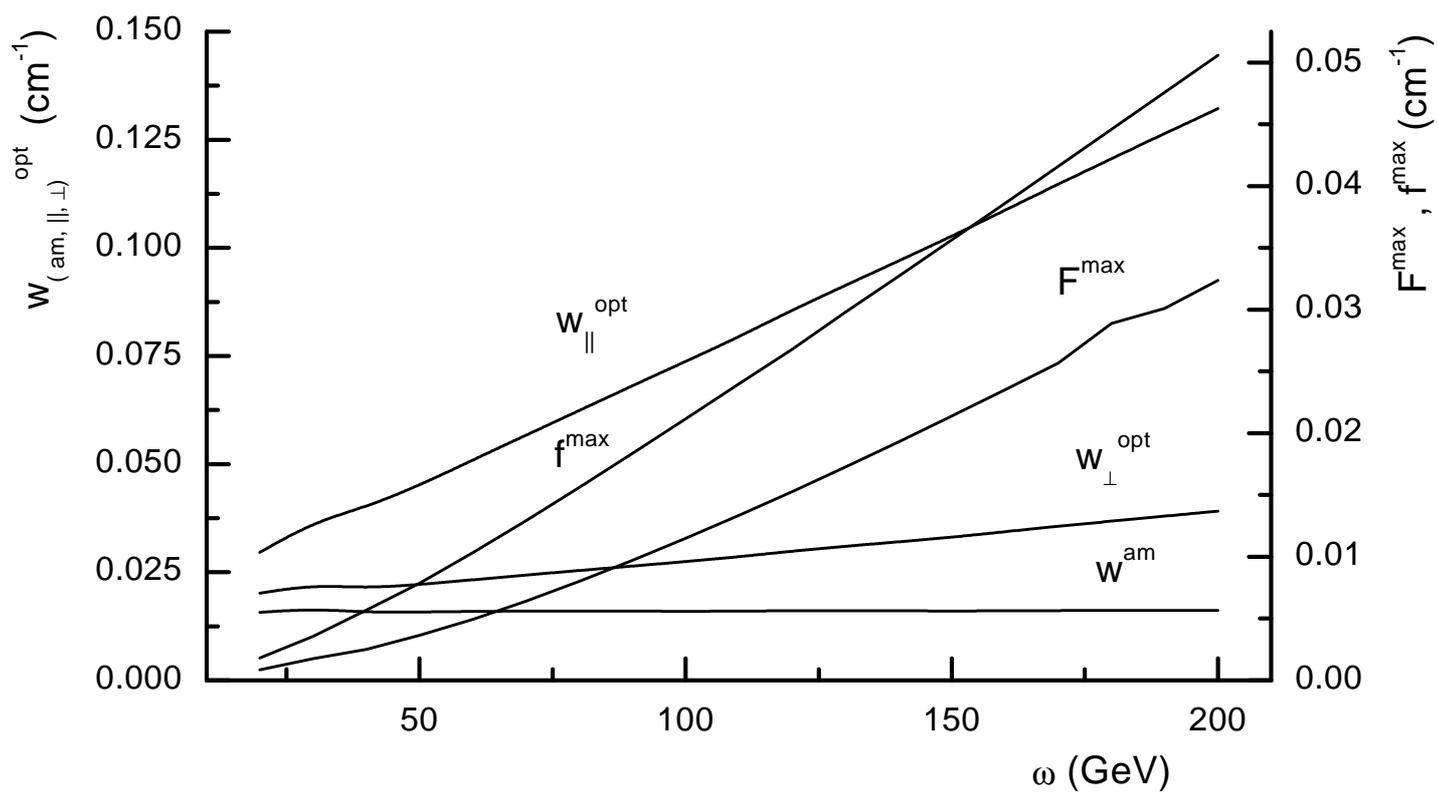
\end{document}